\documentclass{article}
\usepackage{amsmath}
\usepackage{graphicx}
\usepackage{amsfonts}
\usepackage{amssymb}
\usepackage{epsfig}

\begin{document}

\title{Modeling Nonequilibrium Phase Transitions and Critical Behavior 
in Complex Systems}

\author{J. Marro, J.M. Cortes and P.I. Hurtado \\
Instituto Carlos I de F\'{\i}sica Te\'{o}rica y Computacional, \\
Universidad de Granada, 18071-Granada, Espa\~{n}a}

\maketitle

\begin{abstract}
We comment on some recent, yet unpublished results concerning instabilities
in complex systems and their applications. In particular, we briefly describe
main observations during extensive computer simulations of two lattice
nonequilibrium models. One exhibits robust and efficient processes of
pattern recognition under synaptic coherent activity; the second example
exhibits interesting critical behavior and simulates nucleation and spinodal
decomposition processes in driven fluids.


\end{abstract}

\section{Introduction}

Nature may be viewed as a collection of \textit{complex systems} \cite{com}.
Consequently, a principal question is how these systems, which typically
consist of many simple interacting units, develop qualitatively new and
high-level kinds of organization. This is the problem of connecting the
microscopics of constituents with the coherent structures that characterize
organisms and communities. It may often be assumed that the fundamental laws
of physics, such as Hamilton and Maxwell equations, are individual
properties of the units. Still, it is only very rare that the origin and
form of natural phenomena can be inferred from basic laws. What is the
relevance of fundamental physics to predict the weather, to design new
materials and drugs or to understand the origin of life? It is remarkable
that statistical physics recently addressed the problem of connecting
emergent behavior to the constituents' properties in a more indirect
manner, too. That is, main concepts in the theory of phase transitions, such as
correlations, criticality, scale invariance and self-similarity that
characterize the global behavior of the simplest model cases happen to be
ubiquitous in nature.\ This brings many interesting, high-level phenomena to
the attention of physicists, and the study of (nonequilibrium) phase
transitions has consequently been animated \cite{DM}.

As a matter of fact, an important observation in nature is that the complex
systems of interest are often open, and out of a thermodynamic equilibrium
state. Their simplest condition is that of a nonequilibrium steady state.
That is, a constant flux of some quantity (matter, energy,...) is typically
involved and the state is, in general, not determined solely by external
constraints, but depends upon their \textit{history} as well. Under such a
nonequilibrium condition, as the control parameters ---temperature or
potential gradients, or reactant feed rates, for instance--- are varied, the
steady state may become unstable and be replaced by another (or, perhaps, by
a periodic or chaotic state). Nonequilibrium instabilities are attended by
ordering phenomena so analogous to those of equilibrium statistical
mechanics that one may speak of \textit{nonequilibrium phase transitions.}
These are ubiquitous in physics and biology, and have also been described in
the social sciences \cite{DM,nept1,nept2,nept3}.

The simplest examples of nonequilibrium phase transitions occur in lattice
models. The analysis of more realistic situations is presently confronted,
among other problems, with the lack of a general formalism, analogous to
equilibrium statistical mechanics. That is, nonequilibrium dynamics is not
derivable from an energy function. One must actually find time-independent
solutions of master or kinetic equations, which is a formidable task in
practice. Therefore, general theoretical approaches are scarce. It is true
that, for cases in which fluctuations are of minor significance, a
macroscopic description, i.e., a set of partial differential equations is
often preferable to a lattice model, for instance, in predicting a
nonequilibrium phase diagram. However, such macroscopic descriptions imply
mean-field behavior, while lattice models exhibit a range of critical
phenomena and other details which are at least as interesting as in
equilibrium \cite{DM}. The lack of theory also explains that most interesting
information has been gained by means of computer simulations of the lattice
models.

\section{Neural cellular automata that efficiently recognize a pattern}

As a first example of a complex lattice system that exhibits nonequilibrium
phase transitions, let us consider an artificial neural network that was
introduced and studied before \cite{net0}. This consists of a set of $N$
binary \textit{neurons,} $\mathbf{s=\{}s_{\mathbf{x}}=\pm1;\mathbf{x}%
=1,\ldots,N\},$ evolving in time by stochastic equations, 
\begin{eqnarray}
\partial_{t}P_{t}(\mathbf{s},\mathbf{J}) =  p{\displaystyle\sum\limits_{\mathbf{x}}}
[-\varpi_{\mathbf{J}}(s_{\mathbf{x}}\rightarrow-s_{\mathbf{x}})P_{t}(\mathbf{s},
\mathbf{J}) 
+\varpi_{\mathbf{J}}(s_{\mathbf{x}}\rightarrow-s_{\mathbf{x}})P_{t}(\mathbf{s}^
{\mathbf{x}},\mathbf{J})] \nonumber \\
 + (1-p){\displaystyle\sum\limits_{\mathbf{x,y}}}{\displaystyle\sum\limits_{J\mathbf{^
{\prime}}_{xy}}}[-\varpi(J_{xy}\rightarrow J\mathbf{^{\prime}}_{xy})P_{t}(\mathbf{s},
\mathbf{J}) 
+\varpi(J\mathbf{^{\prime}}_{xy}\rightarrow J_{xy})P_{t}(\mathbf{s},\mathbf{J}^{xy})]
\label{E25}
\end{eqnarray}
Here $\mathbf{J}=\left\{ J_{xy}\in\Re;\mathbf{x,y}=1,\ldots,N\right\} $ is
the configuration of synaptic intensities, and $\mathbf{s}^{\mathbf{x}}$ $%
\left( \mathbf{J}^{xy}\right) $ stands for $\mathbf{s}$ $\left( \mathbf{J}%
\right) $ after the change $s_{\mathbf{x}}\rightarrow$ $-s_{\mathbf{x}}$ $%
\left( J_{xy}\rightarrow J\mathbf{^{\prime}}_{xy}\right) $. The function $%
\varpi(J_{xy}\rightarrow J\mathbf{^{\prime}}_{xy})$ is taken independent of
the current $\mathbf{s,}$ and $\varpi_{\mathbf{J}}(s_{\mathbf{x}%
}\rightarrow-s_{\mathbf{x}})=\varphi\left( 2T^{-1}s_{\mathbf{x}}h_{\mathbf{x}%
}\right) ,$ where 
\begin{equation}
h_{\mathbf{x}}=h_{\mathbf{x}}(\mathbf{s},\mathbf{J})=\sum_{\mathbf{y}%
}J_{xy}s_{\mathbf{y}}
\end{equation}
is a local field.

For $p=1,$ (\ref{E25}) reduces to the familiar Hopfield model in which the
neurons evolve in the presence of a set of (frozen) synaptic intensities. It
is assumed that these in some way contain information from a set of $P$
stored patterns, $\mathbf{\xi}=\left\{ \xi_{\mathbf{x}}=\pm1;\mathbf{x}%
=1,\ldots,N\right\} ,$ e.g., the Hebb choice $J_{xy}\propto\sum_{\mu=1}^{P}%
\xi_{\mathbf{x}}^{\mu}\xi_{\mathbf{y}}^{\mu}$ after appropriate
normalization. Under such conditions, the model asymptotically tends to the
equilibrium state for temperature $T$ and energy function, $H=\sum _{\mathbf{%
x}}h_{\mathbf{x}}s_{\mathbf{x}}.$ This state sometimes corresponds to a
configuration closely resembling one of the stored patterns; the system is
therefore said to exhibit \textit{associative memory}. However, this simple
case is not sufficiently efficient for applications; e.g., errors when
recovering a given pattern are large for most values of $N,$ $P$ and $T,$
and the steady state may not be ``pure'' but correspond to a mixture of two
or more stored patterns.

For $p\rightarrow0,$ equation (\ref{E25}) transforms \cite{DM} into 
\begin{equation}
\partial_{t}P_{t}(\mathbf{s})=\sum_{\mathbf{x}}\left[ \varpi(\mathbf{s}^{%
\mathbf{x}};\mathbf{x})P_{t}(\mathbf{s}^{\mathbf{x}})-\varpi(\mathbf{s};%
\mathbf{x})P_{t}(\mathbf{s})\right] ,
\end{equation}
where the transition probability per unit time is the superposition 
\begin{equation}
\varpi(\mathbf{s};\mathbf{x})=\int d\mathbf{J}f(\mathbf{J})\varphi\left[
2T^{-1}s_{\mathbf{x}}h_{\mathbf{x}}\left( \mathbf{s},\mathbf{J}\right) %
\right] .
\end{equation}
For appropriate choices of this superposition, i.e., of functions $f$ and $%
\varphi,$ this system behaves qualitatively differents from the Hopfield
case. That is, it can be shown ---analytically in some cases and, more
generally, by computer simulations--- that a second-order (equilibrium)
phase transition for $p=1$ transforms for $p\rightarrow0$ into a first-order
(nonequilibrium) phase transition. This has some dramatic consequences
concerning the recognition of a given pattern out of a deteriorated image of
it. In particular, for a wide and practically interesting range of $N,$ $P$
and $T,$ mixture states do not occur and the recovery process happens to be
rather robust and accurate \cite{net0}.

This study induced us to investigate a cellular automaton version of the
original model. Firstly, the function $\varphi $ is properly determined;
this choice importantly affects in practice some of the system properties,
e.g., the nature of its phase transitions. The simulation then proceeds by
choosing at random any of the stored patterns, say $\mu ,$ and updating all
the neurons in the lattice assuming the set $J_{xy}=\xi _{\mathbf{x}}^{\mu
}\xi _{\mathbf{y}}^{\mu },$ i.e., the synaptic intensities corresponding to
the selected pattern. Next, this step is repeated again and again. The draw
is performed in such a way 
that the time average for each local $J_{xy}$
gives Hebb's rule, $\left\langle J_{xy}\right\rangle \propto \sum_{\mu
=1}^{P}\xi _{\mathbf{x}}^{\mu }\xi _{\mathbf{y}}^{\mu }$ (or, alternatively,
any other learning rule one may use in the system definition).

%\begin{equation*}
%\FRAME{itbphFU}{3.8086in}{3.2664in}{0in}{\Qcb{\emph{Figure 1}. The time
%evolution of the overlap in our cellular automata (two upper curves) is
%compared here with the Hopfield case (two lower curves). See the text for
%explanations.}}{}{cellular_automata.ps}{\special{language "Scientific
%Word";type "GRAPHIC";maintain-aspect-ratio TRUE;display "USEDEF";valid_file
%"F";width 3.8086in;height 3.2664in;depth 0in;original-width
%7.0024in;original-height 5.9975in;cropleft "0";croptop "1";cropright
%"1";cropbottom "0";filename
%'C:/JMdocs/jm/PAPERS/Aachen/cellular_automata.ps';file-properties "XNPEU";}}
%\end{equation*}

The preliminary results that are available at the time of this writing
reveal that this case exhibits a very robust and efficient process of
pattern recognition. For most parameter values, starting from a perturbed
pattern, the system rapidly transforms that configuration into the stored
pattern that is closest to it. Figure 1 shows the time evolution of the
overlap between the actual state of the system and one of the stored
patterns at indicated temperature. This is in units of the respective model critical
temperature, either $T_{C}^{\ast }=0.1$ and $T_{C}^{\ast \ast }=1$ for our
system and for the Hopfield case, respectively. \emph{Type 1} here refers to the
case in which the system stores $P=150$ patterns whose sites are generated
completely at random, so that each site is independent of the others; 
\emph{type 2} is for $P=90$ stored patterns generated using the logistic map
in the chaotic region of its parameter space, so that some correlation exits
between sites. In both cases, our algorithm rapidly detects the pattern which is 
closest to the initial state. This behavior holds essentially for other values of the 
parameters.

The reason behind the \textit{good }properties of our system seems to be that,
for appropriate dynamics, the actual state only evolves noticeably
at steps in which synapses correspond to the selected pattern \cite{nets}.
This behavior opens the model to a wide range of possible applications.

\section{Spinodal decomposition and criticality in driven fluids}

The driven lattice gas (DLG) is a $d-$dimensional ($d=2$ in the following)
lattice gas at temperature $T$ in which transitions in (against) one of the
principal lattice directions ---say $\parallel,$ to be referred to as the 
\textit{field direction---} are favored (unfavored). For periodic boundary
conditions, this induces a net current of particles along the field
direction. At high $T,$ the system is in a disordered state while, for
half-filled lattices (the only case of interest in this paper), there is a
second-order (nonequilibrium) critical point, below which the DLG segregates
showing anisotropic, stripe-like configurations parallel to the field \cite{DM}.

Establishing the universality class of the DLG is a main issue not only
concerning a better understanding of these model properties but also much
more generally, in relation to the theory of nonequilibrium phase
transitions and critical phenomena. In fact, the DLG is recognized as one of
the more intriguing model examples of nonequilibrium phenomena. 

Recent field theory \cite{DLG,field} motivated performing new and extensive computer
simulations. These focused on the case of an ``infinite'' drive (particles
along the field direction are not allowed to go backwards) for both \textit{%
large} squares and rectangular $L_{\parallel }\times L_{\perp }$ lattices of
different, appropriate sizes. It has thus been demonstrated numerically that
the DLG belongs to the same universality class as a lattice gas under a
randomly fluctuating field,\cite{DM} so that (using the renormalization
group jargon) the particle current ---which does not occur in the latter---
is not a \textit{relevant }feature of the DLG. Main critical exponents
follow for both cases as $\beta =0.33(1),$ $\nu _{\parallel }\simeq 1.25$
and $\nu _{\parallel }\simeq 2\nu _{\perp }.$ These important results (that
have more recently been confirmed independently \cite{alba}) can be
interpreted within the context of the existing field theory \cite{DLG,field}.
It should be noted however, that the present form of this theory does not fit well the results of
further numerical investigation of the same model. That is, studying small
values of the field suggests that the DLG has no relation to the
equilibrium lattice gas ---so that one cannot go perturbatively from the
latter to the case of small fields. There is some indication that the DLG
belongs to the same universality class with $\beta <1/2$ for any value of
the field as long as the configurations are stripped \cite{DLGmc}. If this is
confirmed, the chances are that the DLG will again attract considerable
attention during the coming years.

%\begin{equation*}
%\FRAME{itbphFU}{3.8052in}{3.1349in}{0in}{\Qcb{\emph{Figure 2. }Scaling of $%
%S(k,t)$ for different times between $10^{5}$ MCS and $10^{6}$ MCS in a 128$%
%\times $128 system at low temperature, $T=0.57T_{C}.$ Two lines of slope $-2$
%and $-3,$ respectively, are indicated. The inset depicts the width stripe $%
%\ell (t)$ versus $t^{1/3}$ for the same system.}}{}{joaquin-alemania.eps}{%
%\special{language "Scientific Word";type "GRAPHIC";maintain-aspect-ratio
%TRUE;display "USEDEF";valid_file "F";width 3.8052in;height 3.1349in;depth
%0in;original-width 6.5138in;original-height 5.3558in;cropleft "0";croptop
%"1";cropright "1";cropbottom "0";filename
%'C:/JMdocs/jm/PAPERS/Aachen/joaquin-Alemania.eps';file-properties "XNPEU";}}
%\end{equation*}

In fact, in addition to the above issue on criticality, there are further
interesting questions concerning this model. One is the nature of its
kinetic behavior as a configuration evolves from a disordered state to the
stripped one. Extending the arguments first checked for systems that evolve
towards equilibrium,\cite{mlk} one should perhaps expect \textit{%
self-similarity} with time of the structure function. That is, as the system
undergoes nucleation and then cluster coagulation according to a sort of
spinodal decomposition, even though this is strongly anisotropic and will
eventually lead to a nonequilibrium steady state, it seems reasonable to
assume the existence of a unique relevant length. One should expect this to
be the length that characterizes the (transverse) clustering process in the
system, say $\ell\left( t\right) $. Consequently, quantities changing with
time should not depend explicitly on the time variable but only through $\ell\left(
t\right) .$ For example, the structure function $S\left( k,t\right) $ is
expected to depict a scaled form ${\tilde s}(k'),$ independent of $t,$ when plotted
accordingly,

\begin{equation}
S(k,t) \propto \ell(t) {\tilde s} (k \ell(t))
\end{equation}

This has recently been confirmed using the width of the stripes
as the relevant length, as shown in Fig. 2. Moreover, we found that 
${\tilde s} (k) \sim k^{-2}$ for large values of $k$, which is the generalization of 
Porod's law to DLG. This power law behavior, as compared to the $k^{-3}$ tail observed 
in bidimensional equilibrium binary mixtures, reflects the fact that coarsening in DLG is 
effectively a unidimensional process which takes place in the direction perpendicular to the field. 

It has also been shown that $\ell\left( t\right) \sim t^{1/3},$ in general, as in standard 
spinodal decomposition,\cite{DLGkinetics} though the mechanisms leading to this behavior seem to be
different. More specifically, we found that $\ell(t) = a(t/L_{\parallel})^{1/3} + b$ for long 
enough times, where $a$ and $b$ are constants which depend on temperature, while for intermediate 
times we observe $\ell(t) \sim (t/L_{\parallel})^{1/4}$ (see Fig. 3). 

The analysis of all the above results on the dynamics of DLG allows us to conclude that the 
particle current does not play any important role in the late stage
coarsening of DLG, the anisotropy being the relevant ingredient present in this process.

These facts motivate investigating experimentally spinodal
decomposition in samples under nonequilibrium anisotropic conditions, e.g., scattering
studies of fluids under shear.\bigskip

\noindent\textbf{Acknowledgments. }The unpublished work partially reported
in this paper was performed in collaboration with A. Achahbar, E. Albano,
P.L. Garrido, M.A. Mu\~{n}oz and J.J. Torres; a proper account of it is
given elsewhere \cite{nets,DLG,DLGmc,DLGkinetics}. We also acknowledge
financial support under the Spanish projects DGESEIC PB97-0842 and SEPCYT
BFM2001-2841.

\newpage

%\noindent \textbf{Figure Captions}

%\begin{itemize}

%\item  {\bf Figure 1: } The time evolution of the overlap in our cellular automata (two upper curves) is
%compared here with the Hopfield case (two lower curves). Note that, on the time scale of the experiment, 
%only our system depicts a clear tendency towards saturation. See the text for more explanations.

%\item {\bf Figure 2:} Scaling of $S(k,t)$ for different times between $10^{5}$ MCS and $10^{6}$ MCS 
%in a $128 \times 128$ system at low temperature, $T=0.57T_{C}.$ Two lines of slope $-2$
%and $-3,$ are indicated. 

%\item {\bf Figure 3:}  In this figure we show $\ell(t)$ versus $t^{\alpha}$ for both
%$L_{\perp} \times L_{\parallel} = 64 \times 64$ $(a)$ and $256 \times 64$ $(b)$, 
%and where $\alpha = 1/4$ and $1/3$ has been used, respectively. Notice that both plots 
%follow a lineal law of the form $\ell(t) = a t^{\alpha} + b$. $1/4$-behavior dominates the full 
%evolution of the smaller system, while the larger one reaches the long time limit, thus showing 
%$1/3$-behavior. In the insets we show $\chi ^2 (\alpha)$ for three different measures of the mean 
%stripe width $\ell(t)$. Its minimum signals the experimental value of the growth exponent $\alpha$. 

%\end{itemize}

%\newpage

%%  figure 1
\begin{figure}[t]
\centerline{
\epsfxsize=3in
\epsfysize=2.5in
\epsfbox{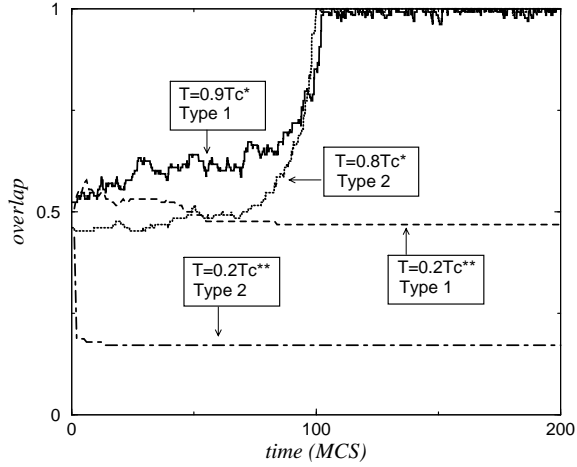}
}
\caption{The time evolution of the overlap in our cellular automata (two upper curves) is
compared here with the Hopfield case (two lower curves). Note that, on the time scale of the experiment, 
only our system depicts a clear tendency towards saturation. See the text for more explanations.  }
\label{neuronas}
\end{figure}

%%  figure 2

\begin{figure}
\centerline{
\epsfxsize=3in
\epsfysize=2.5in
\epsfbox{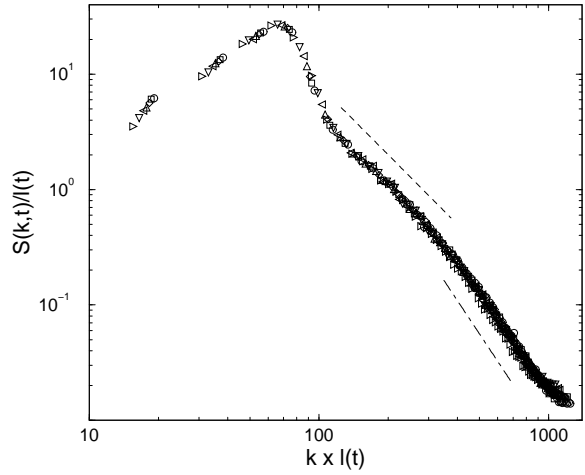}
}
\caption{Scaling of $S(k,t)$ for different times between $10^{5}$ MCS and $10^{6}$ MCS 
in a $128 \times 128$ system at low temperature, $T=0.57T_{C}.$ Two lines of slope $-2$
and $-3,$ are indicated. }
\label{colapso}
\end{figure}

%% figure 3  

\begin{figure}
\centerline{
\psfig{file=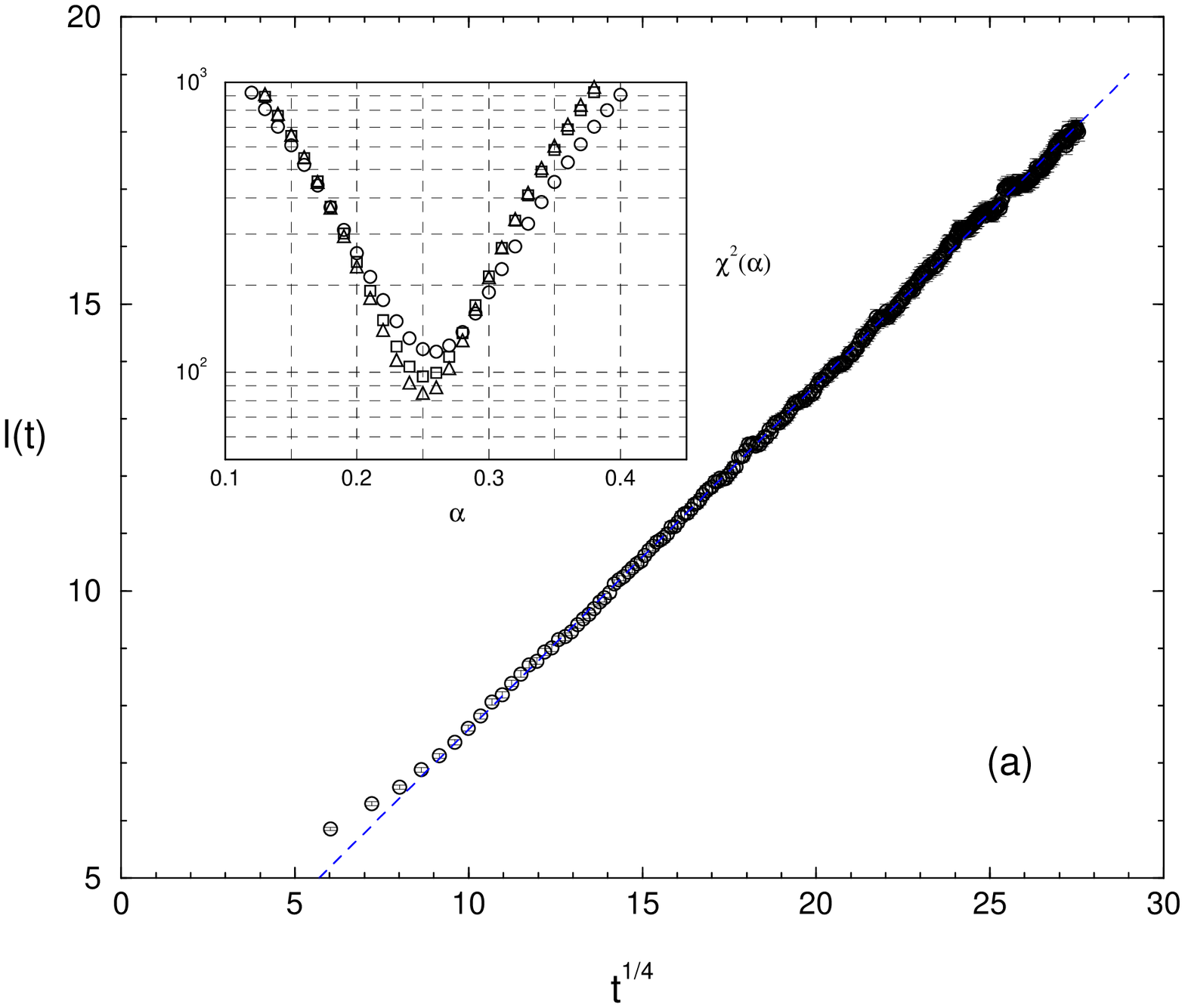,width=7.5cm,angle=-90}
%\epsfxsize=3in
%\epsfysize=3in
%\epsfbox{fig3a_id_723.eps}
}
\centerline{
\psfig{file=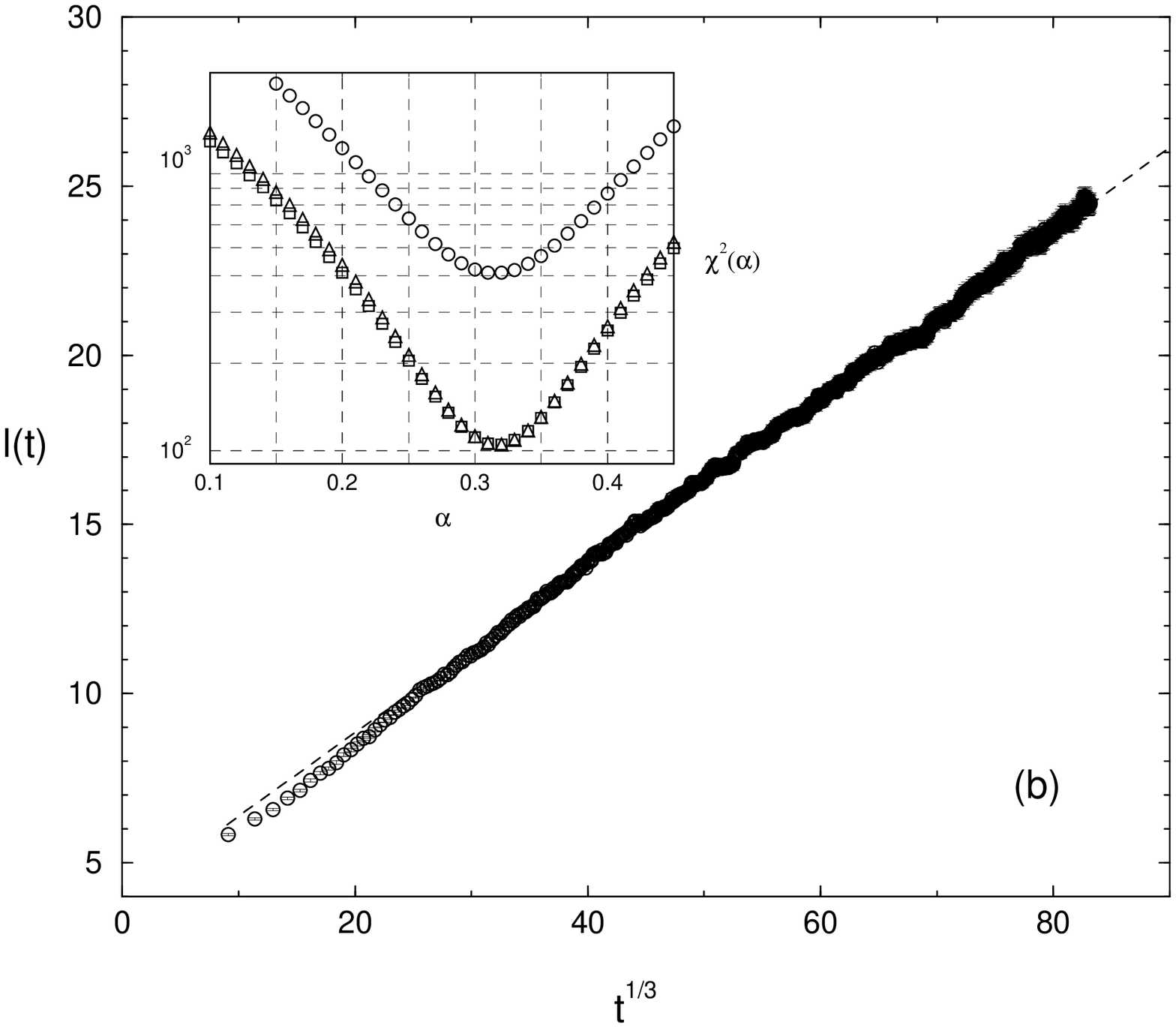,width=7.5cm,angle=-90}
%\epsfxsize=3in
%\epsfysize=3in
%\epsfbox{fig3b_id_723.eps}
}
\caption{In this figure we show $\ell(t)$ versus $t^{\alpha}$ for both
$L_{\perp} \times L_{\parallel} = 64 \times 64$ $(a)$ and $256 \times 64$ $(b)$, 
and where $\alpha = 1/4$ and $1/3$ has been used, respectively. Notice that both plots 
follow a lineal law of the form $\ell(t) = a t^{\alpha} + b$. $1/4$-behavior dominates the full 
evolution of the smaller system, while the larger one reaches the long time limit, thus showing 
$1/3$-behavior. In the insets we show $\chi ^2 (\alpha)$ for three different measures of the mean 
stripe width $\ell(t)$. Its minimum signals the experimental value of the growth exponent $\alpha$.  }
\label{ancho_chi2_64}
\end{figure}

\end{document}